\documentclass[aps,10pt,prb,twocolumn,showpacs,floatfix,longbibliography]{revtex4-1} 
\usepackage{amsfonts}
\usepackage{graphicx}
\usepackage{amsmath}
\usepackage{times}
\usepackage{amssymb}
\usepackage{color}
\usepackage[colorlinks,bookmarks=false,citecolor=blue,linkcolor=red,urlcolor=blue]{hyperref}

\begin{document}

\title{A single-layer tensor-network study of kagome Heisenberg model with chiral interaction}
\author{R. Haghshenas$^1$}
\author{Shou-Shu Gong$^2$}
\author{D. N. Sheng$^1$}
\affiliation{$^1$Department of Physics and Astronomy, California State University, Northridge, California 91330, USA\\ $^2$Department of Physics, Beihang University, Beijing, 100191, China}

\begin{abstract}
We study the antiferromagnetic kagome Heisenberg model with additional scalar-chiral interaction by using the infinite projected entangled-pair state (iPEPS) ansatz. We discuss in detail the implementation of optimization algorithm in the framework of the single-layer tensor network based on the corner-transfer matrix technique. Our benchmark based on the full-update algorithm shows that the single-layer algorithm is stable, which leads to the same level of accuracy as the double-layer ansatz but with much less computation time. We further apply this algorithm to study the nature of the kagome Heisenberg model with a scalar-chiral interaction by computing the bond dimension scaling of magnetization, bond energy difference, chiral order parameter and correlation length. In particular, we find that for strong chiral coupling the correlation length, which is extracted from the transfer matrix, saturates to a finite value for large bond dimension, representing a gapped spin-liquid state. Further comparison with density matrix renormalization group results supports that our iPEPS faithfully represents the time-reversal symmetry breaking chiral state. Our iPEPS simulation results shed new light on constructing PEPS for describing gapped chiral topological states.
\end{abstract}
\pacs{75.40.Mg, 75.10.Jm, 75.10.Kt,  02.70.-c}
\maketitle

%%%%%%%%%%% 
\section{Introduction}

Understanding low-energy physics of geometrically-frustrated quantum spin systems has been one of the main active areas of research in modern condensed matter physics in last decades~\cite{Balents:2010,savary:2016}. Generally, in a frustrated system, quantum fluctuations are enhanced as the local energetic interactions cannot be simultaneously minimized. This property may lead the system into novel liquid-like phases in which magnetic long-range order is absent~\cite{read:1989, read:1990, Laughlin:1981, Yang:1993, Anderson:1973, Wen:1990, Read:1991}. Specifically, spin-liquid phases exhibit exotic characteristics including long-range entanglement~\cite{Chen:2010} and anyonic braiding statistics of quasiparticle excitations~\cite{Wen:1990, Wen:2007}, making them an important playground to realize quantum computation~\cite{kitaev2006}. Besides quantum spin liquids, fractional quantum Hall effects~\cite{Stormer:1983} can also realize topological ordered states, which have been discovered in two-dimensional correlated electron gas subject to a magnetic field. Such states are characterized by Abelian or non-Abelian fractional statistics, the bulk gap (insulator) and gapless chiral edge modes with universal spectra~\cite{Wen:1991, Wen:1992}.

Among various frustrated magnetic systems, the spin-$1/2$ kagome Heisenberg model with nearest-neighbor interaction has been studied extensively as a promising candidate to host a spin-liquid state in frustrated quantum magnetism~\cite{Sachdev:1992, Hastings:2000, Singh:2007}. The most promising numerical studies, i.e. density matrix renormalization group (DMRG) and tensor-network-based methods, in the last few years, have predicted a possible spin-liquid state, although its nature is still controversial~\cite{Yan:2011, jiang2012, depenbrock2012, He:2017, Liao:2017}. The gapped $\mathcal{Z}_2$ spin liquid and gapless $U(1)$ Dirac spin liquid~\cite{ran2007, iqbal2011, iqbal2013, iqbal2014} are the main candidates for this phase, referred as kagome spin liquid in this paper. Experimentally, the Herbertsmithite ZnCu$_3$(OH)$_6$Cl$_2$ has been considered as one of the leading candidate materials for the $S = 1/2$ kagome Heisenberg model. In particular, the inelastic neutron scattering~\cite{Helton:2007} and nuclear magnetic resonance measurements~\cite{Fu:2015} on Herbertsmithite provide strong evidence to support a spin-liquid ground state, but it is unclear whether the low-energy excitations are gapped or gapless.

One of the well established spin liquids in microscopic models is the chiral spin liquid, which was first introduced by Kalmeyer and Laughlin on a two-dimensional lattice~\cite{Kalmeyer:1987, Wen:1989}. This state actually represents a lattice version of the fractional quantum Hall state in which the time-reversal and parity symmetries are broken, relating to non-zero scalar chirality term $\textbf{S}_{i} \cdot (\textbf{S}_{j} \times \textbf{S}_{k})$ mimicking orbital magnetic field. For this exotic phase of matter, the sharp property is the emergence of semionic fractional statistics---exchanging two semions, resulting in multiplying the wave function by a complex phase $i$---and their striking topologically protected chiral edge physics described by a chiral conformal field theory~\cite{Halperin:1982, Wen:1990, Yang:1993}. It has been shown recently that some local Hamiltonians, on the kagome and square lattices, could host such chiral spin-liquid states~\cite{Gong:2014,Wietek:2015, He:2014, Poilblanc:2017, Poilblanc:2018}. Based on DMRG studies~\cite{Bauer:2014}, a chiral spin-liquid phase was established on the kagome Heisenberg model with an additional scalar-chiral interaction. It is shown that this chiral spin-liquid state inherits universal features of the $\nu=\frac{1}{2}$ Laughlin state, which remains stable in a wide range of coupling parameter~\cite{Bauer:2014}.

Theoretically, a well-controlled analytical tool for studying strongly correlated systems is rare, and large-scale quantum Monte Carlo simulation suffers from sign problem. In recent years, the promising two-dimensional tensor-network ansatz infinite projected entangled-pair state (iPEPS) has become a powerful method to study such systems~\cite{Orus:2014, Verstraete:2008}. The iPEPS provides a variational ansatz, which approximates the ground-state wavefunction of a two-dimensional system directly in the thermodynamic limit. 
Its success lies in the fact that it can capture the entanglement area law---which is the case for the ground states of a wide range of two-dimensional local Hamiltonians. The only essential parameter which controls the accuracy of the ansatz is the so-called bond dimension $D$. In order to obtain highly accurate results, one should use a novel optimization scheme to access large bond dimension and to perform reliable bond-dimension scaling. iPEPS has been shown successful to study challenging problems of interacting fermions (including $t$-$J$ and Hubbard models)~\cite{Corboz:2011, Corboz:2014, Corboz:2016} and frustrated spin systems~\cite{Bauer:2012, Corboz:2013, Osorio:2014, Haghshenas:2018May, Haghshenas:2018, Jahromi:2018}. Besides model simulation, PEPS can also be constructed to strictly describe novel quantum states. While non-chiral topologically ordered states have been accurately characterized in PEPS formalism~\cite{Levin:2005, schuch2012, poilblanc2012}, the chiral topological PEPS that have been proposed so far are always gapless~\cite{dubail2015, wahl2013, yang2015, poilblanc2015}. Although the topological properties such as chiral edge mode and topological entanglement entropy have been found in the chiral PEPS, the correlation length is diverging, which indicates gapless bulk excitations~\cite{dubail2015, wahl2013, yang2015, poilblanc2015}. Therefore, how to construct a gapped chiral topological PEPS is still a challenging question. Meanwhile, model simulations for correlated systems to identify gapped chiral topological states based on PEPS are rare.

In this paper, our main objectives include $(1)$ to examine an efficient optimization technique in the framework of the single-layer tensor network proposed in Ref.~\onlinecite{Xie:2017}, which enables us to further increase bond dimension $D$ in the iPEPS simulation without losing accuracy; (2) to investigate the bulk properties of the kagome Heisenberg model with additional scalar-chiral interaction $J_{ch} \textbf{S}_{i} \cdot (\textbf{S}_{j} \times \textbf{S}_{k})$ ($J_{ch}$ is the strength of the chiral coupling). We present details on implementation of the so-called full-update optimization method used in the iPEPS ansatz in a single-layer tensor network by using corner transfer matrix (CTM) method~\cite{SL:note}. We show how to accomplish this algorithm with a computational cost that scales with bond dimension as $\mathcal{O}(D^9)$. In addition, we provide benchmark data for the square Heisenberg model and compare it with previous well-known results. We demonstrate that this algorithm is stable and we can obtain accurate results that are similar to the conventional double-layer algorithm but with less computation time.

By using our scheme to study the kagome Heisenberg model with chiral interaction, we show that the iPEPS ansatz provides quite competitive variational energy compared to the energy from DMRG. We calculate the local order parameters including magnetization, local bond energy difference (detecting lattice symmetry breaking) and chiral order to study the nature of the system. Specifically, for $J_{ch} > 0.2$ (we take the Heisenberg coupling as energy scale) where the system has been shown in the gapped chiral spin-liquid phase in DMRG calculation~\cite{Bauer:2014}, the magnetization and lattice order parameters vanish quite fast with growing bond dimension. By studying the correlation length extracted from the transfer matrix, we find finite correlation length in large bond dimension $D > 12$ and thus identify the ground state as a gapped chiral spin liquid. Furthermore, by checking the coupling dependence of the scalar-chiral order, we estimate the quantum phase transition between the chiral spin liquid and the kagome spin-liquid phase at $J_{ch} \approx 0.14$ based on our studied bond dimension. Some difference between iPEPS and DMRG results at small $J_{ch}$ will be discussed. This chiral spin-liquid phase with finite correlation length found in our study not only shows an example of gapped chiral state identified by iPEPS simulation, but also sheds new light on PEPS construction of gapped chiral topological states in the future study.

This paper is organized as follows: In Sec.~\ref{subsec:iPEPS-ansatz}, we briefly review iPEPS ansatz by discussing the contraction scheme and the full-update optimization method in a double-layer tensor-network framework. We further show in detail how to do contractions based on corner transfer matrix (Sec.~\ref{subsec:A-single-layer-corner-transfer}) and full-update optimization (Sec.~\ref{Sec:single-layer-full-update}) in a single-layer tensor-network framework. We present benchmark results on square Heisenberg model in Sec.~\ref{Sec:Benchmark}. In Sec.~\ref{sec:chiral}, we introduce the kagome Heisenberg model with an additional chiral interaction and present our iPEPS results of local order parameters and correlation length extracted from transfer matrix. In Sec.~\ref{Sec:CONCLUSION} we summarize and discuss our results.

\section{iPEPS ansatz in a single-layer tensor-network framework}
\label{sec:iPEPS-ansatz}

\subsection{iPEPS ansatz}
\label{subsec:iPEPS-ansatz}

An iPEPS is a tensor-network state defined by a set of local interconnected tensors to efficiently describe the ground-state wave function of a two-dimensional quantum system. The tensors, shown by $\{A\}$, are connected to each other by the so-called virtual bonds to form a geometrical pattern similar to the actual lattice. As depicted in Fig.~\ref{fig:ipeps}(a), we have shown an uniform iPEPS wavefunction $|\Psi \rangle$ on the square lattice (respecting one-site translational invariant symmetry)
\begin{equation*}
|\Psi \rangle = \sum_{i}   \mathcal{F}(\{A_i\}) |\cdots i \cdots \rangle
\end{equation*}
where $\mathcal{F}$ stands for tensor contraction (of virtual bonds) and index $i$ represents the physical Hilbert space with dimension $d$---i.e. $i=\{0,\cdots , d-1\}$. The open bonds are called physical bonds representing spins. The virtual bonds with the so-called bond dimension $D$ control the amount of entanglement in the wavefunction. The iPEPS could be used as a variational ansatz to approximate the ground-state wavefunction of a two-dimensional system: the main goal is to optimize the tensors in order to minimize the variational ground-state energy. The bond dimension $D$ controls the number of variational parameters and hence the accuracy of ansatz. In order to accomplish an iPEPS algorithm, one needs to do two key tasks: $(i)$ finding an efficient algorithm to contract a two-dimensional tensor-network $\mathcal{E}$ produced by, e.g., the scalar product $\langle  \Psi| \Psi \rangle$ and $(ii)$ an optimization method to determine tensors $\{A\}$, so that the variational ground-state energy becomes minimum. As it has been discussed extensively in literature~\cite{Verstraete:2008,Orus:2014,Phien:2015, Corboz:2010:April}, both tasks could be done in the framework of corner transfer matrix renormalization group approach with computational cost $\mathcal{O}(D^{12})$. Our goal is to reformulate these algorithms into a more efficient approach~\cite{Xie:2017} with reduced computational cost $\mathcal{O}(D^9)$ and compare its accuracy with that of previous approaches.

\subsection{A single-layer corner transfer matrix approach}
\label{subsec:A-single-layer-corner-transfer}

\begin{figure}
\begin{center}
\includegraphics[width=1.0 \linewidth]{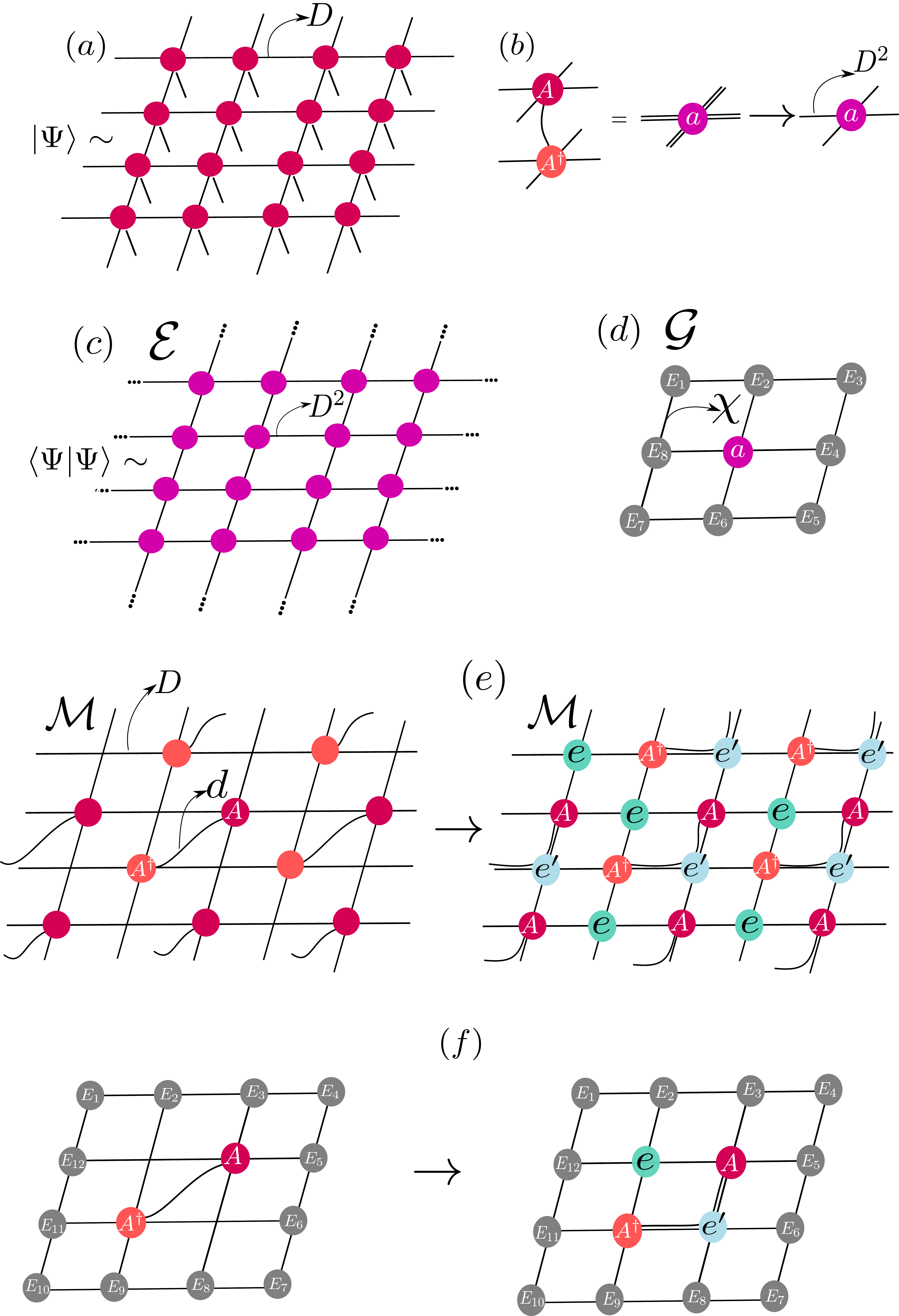} 
  \caption{(Color online) (a) Tensor-network representation of the iPEPS $|\Psi \rangle$, made from local five-rank tensors $\{A\}$ (solid circles). The virtual bonds have bond dimension $D$. (b) The tensor $a$ is defined by contraction on physical index of tensors $A$ and $A^{\dagger}$. In tensor $a$, virtual bonds have bond dimension $D^2$. (c) the scalar product $\langle \Psi| \Psi \rangle$ is obtained by contracting an infinite \emph{double-layer} tensor-network $\mathcal{E}$. (d) A finite tensor-network $\mathcal{G}$, made from environment tenors $E_1, \cdots, E_8$ which are obtained by CTM approach, to approximate tensor-network $\mathcal{E}$. The boundary bond dimension $\chi$ controls accuracy of this approximation. (e) The \emph{one-layer} tensor-network $\mathcal{M}$ represents the scalar product $\langle \Psi| \Psi \rangle$, obtained without tracing over physical index as before. The right figure, representing a $2\times 2$ unit cell tensor network, is driven from left one by reshaping physical bonds and inserting identity tensors $e$ and $e'$ for crossing bonds.  (f) The tensor-network $\mathcal{M}$, made from tensors $\{A, A^{\dagger}, e, e'\}$, could be approximated by using a standard CTM approach adapted to a $2\times 2$ unit cell.}
  \label{fig:ipeps}
\end{center}
\end{figure}

In order to calculate, e.g., the scalar product $\langle  \Psi| \Psi \rangle$, the standard approach is to first trace over physical index of tensors $A$ and $A^{\dagger}$ as depicted in Fig.~\ref{fig:ipeps}(b) to make a reduced tensor $a$. We could then fuse virtual bonds together to make one with larger bond dimension $D^{2}$---we finally end up with a four-rank tensor $a$ (shown by violet circles) with bond dimension $D^2$. As shown in Fig.~\ref{fig:ipeps}(c), the scalar product $\langle \Psi| \Psi \rangle$ is obtained by contracting an infinite \emph{double-layer} tensor-network $\mathcal{E}$, consisting of the reduced tensors $a$ with bond dimension $D^2$. The word \emph{double-layer} is pointing to this fact that tensor-network $\mathcal{E}$ is obtained from compressing two one-layer tensor-networks $|\Psi \rangle$ and $\langle \Psi |$. The double-layer tensor-network $\mathcal{E}$ could be approximated by a smaller finite tensor-network $\mathcal{G}$ based on CTM approach, as shown in Fig.~\ref{fig:ipeps}(d). The corner tensors ${E_1,E_4,E_7,E_{10}}$ and edge tensors ${E_2,E_4,E_6,E_8}$ respectively represent a quadrant and a half row/column of tensor network $\mathcal{E}$.  The bond dimension of environment tensors is denoted by $\chi$ controlling the accuracy of the CTM approach. Generally speaking, to approximate well $\mathcal{E}$ by $\mathcal{G}$, one expects to set $\chi \sim D^{2}$. Thus, the leading cost of carrying out this algorithm would be $\mathcal{O}(D^{12})$ in computational time.

A more efficient strategy, as discussed in Ref.~\onlinecite{Xie:2017}, is to resketch tensor-network representation of $\langle \Psi| \Psi \rangle$ in a single-layer picture without tracing over physical index as shown in Fig.~\ref{fig:ipeps}(e, Left). In order to do that we just need to shift tensors (either in $|\Psi \rangle$ or $\langle \Psi |$) along the $x$-axis (or similarly $y$-axis) by half of a unit cell and express whole infinite tensor network $\langle \Psi| \Psi \rangle$ in a single layer. We denote this tensor network by $\mathcal{M}$, which includes bonds with bond dimension either $D$ or $d$---note the double-layer tensor-network $\mathcal{E}$ has a larger bond dimension $D^2$. This network $\mathcal{M}$ could be approximated in the framework of CTM approach as follows~\cite{SL:note}: we replace the crossing lines with identity tensors $e$ and $e'$ and then relocate the physical bonds by absorbing them into tensors $e'$ as shown in Fig.~\ref{fig:ipeps}(e, Right)---tensors $e$ and $e'$ actually are trivial crossing lines. We finally end up with a $2\times 2$ unit cell tensor-network $\mathcal{M}$, including tensor $A$, $A^{\dagger}$, $e$ and $e'$ which could be handled with a standard CTM approach. We notice that the largest bond dimension in tensor-network $\mathcal{M}$ is $Dd$, which is much smaller than $D^2$, as $d$ is typically small. The infinite tensor-network $\mathcal{M}$ could be easily approximated by an effective finite one specified by the environment tensors $E_1,\cdots, E_6$, as shown in Fig.~\ref{fig:ipeps}(f). The major advantage of representing the tensor-network $\mathcal{E}$ in a single layer is that the leading computation cost scales as $\mathcal{O}(d^{3}D^3\chi^3)$. If we set $\chi=D^{2}$ that would become $\mathcal{O}(D^{9})$ ($d$ is also small) which is much smaller than $\mathcal{O}(D^{12})$. In the iTEBD scheme of this single-layer tensor network, the computation cost also scales as $\mathcal{O}(D^9)$~\cite{Xie:2017}, the same efficiency as the CTM method.

We notice that in the single-layer tensor-network approach, to handle contraction of scalar product $\langle \Psi| \Psi \rangle$, we need to enlarge original unit-cell of iPEPS $|\Psi \rangle$: for example in Fig.~\ref{fig:ipeps}(c), in double-layer tensor-network approach we end up with a $1\times 1$ unit-cell tensor-network $\mathcal{E}$, while in order to handle the same contraction in single-layer tensor-network approach, a $2 \times 2$ unit cell tensor network $\mathcal{M}$ is needed. Specifically, in general, to perform contraction of a $L_x \times L_y$ unit-cell iPEPS $|\Psi \rangle$ in a single-layer tensor-network approach, an enlarged tensor network with $2L_x\times 2L_y$ unit cell should be used. To treat arbitrary unit cells, one could use the approach presented in Ref.~\onlinecite{Corboz:2011}.

 %A discussion about bond dimension should be added...
%%%%%%%%%%%%%%%%%%%%%%%%%%%%%%%%%%Fig. 1%%%%%%%%%%%%%%%%%%%%%%%%%%%%%%%%%%%%%%%%%%
 %%%%%%%%%%%%%%%%%%%%%%%%%%%%%%%%%%%%%%%%%%%%%%%%%%%%%%%

\subsection{A single-layer full-update simulation}
\label{Sec:single-layer-full-update}
The usual optimization method, used to minimize the ground-state energy of iPEPS ansatz, is the so-called full-update scheme. In this scheme, the local tensors are optimized by using the imaginary-time evolution in the class of the iPEPS ansatz. Our goal here is to discuss how this scheme could be adapted in a single-layer tensor-network framework, called \emph{single-layer full update}. We aim to obtain the ground state of a local Hamiltonian $H$, containing only nearest-neighbor interactions $\sum_{i} h_i$ by using the imaginary-time evolution. In order to accomplish this task the imaginary time-evolution operator $e^{-\tau H}$ should be applied to an initial iPEPS state in $\tau \rightarrow \infty$ limit. There are two main obstacles to accomplish this approach: $(i)$ to efficiently express $e^{-\tau H}$ and $(ii)$ to design a truncation procedure as applying time-evolution operator would exponentially increase the bond dimension. We use a first-order Trotter-Suzuki decomposition \cite{Suzuki:1990}, while keeping $\tau$ quite small to approximate the time-evolution operator as follows 
\begin{equation*}
e^{-\tau H} \sim e^{-\tau H_r} e^{-\tau H_l} e^{- \tau H_u} e^{-\tau H_d}+\mathcal{O}(\tau^{2}), \quad \tau << 1, 
\end{equation*}
where indices $\{r, l, u, d\}$ stand for the links labeled by $\{\text{right, left, up, down}\}$. We notice that, e.g., $H_{r}$ only includes the operator terms which commute with each other, i.e. $e^{-\tau H_r}=\otimes_r e^{-\tau h_{r}}$. Thus, we need to evolve an initial iPEPS state $|\psi_{i} \rangle$ with bond dimension $D$ given by
\begin{equation*}
|\psi_{i+1} \rangle \sim e^{-\tau h_r} |\psi_{i} \rangle,
 \end{equation*}
where $|\psi_{i+1}\rangle$ at each step $i$ is effectively represented by an updated iPEPS $|\psi_{i+1} \rangle$ with the same bond dimension $D$. As at each step, the time-evolution operator increases the bond dimension, so a truncation procedure should be applied to prevent exponential growth of parameters. We utilize applications of positive approximant~\cite{Phien:2015, Lubasch:2014} and reduced tensors~\cite{Corboz:2010:April} in an iterative way to handle the truncation algorithm.

%%%%%%%%%%%%%%%%Fig. 2%%%%%%%%%%%%%%%%%%
\begin{figure}
\begin{center}
\includegraphics[width=1.0 \linewidth]{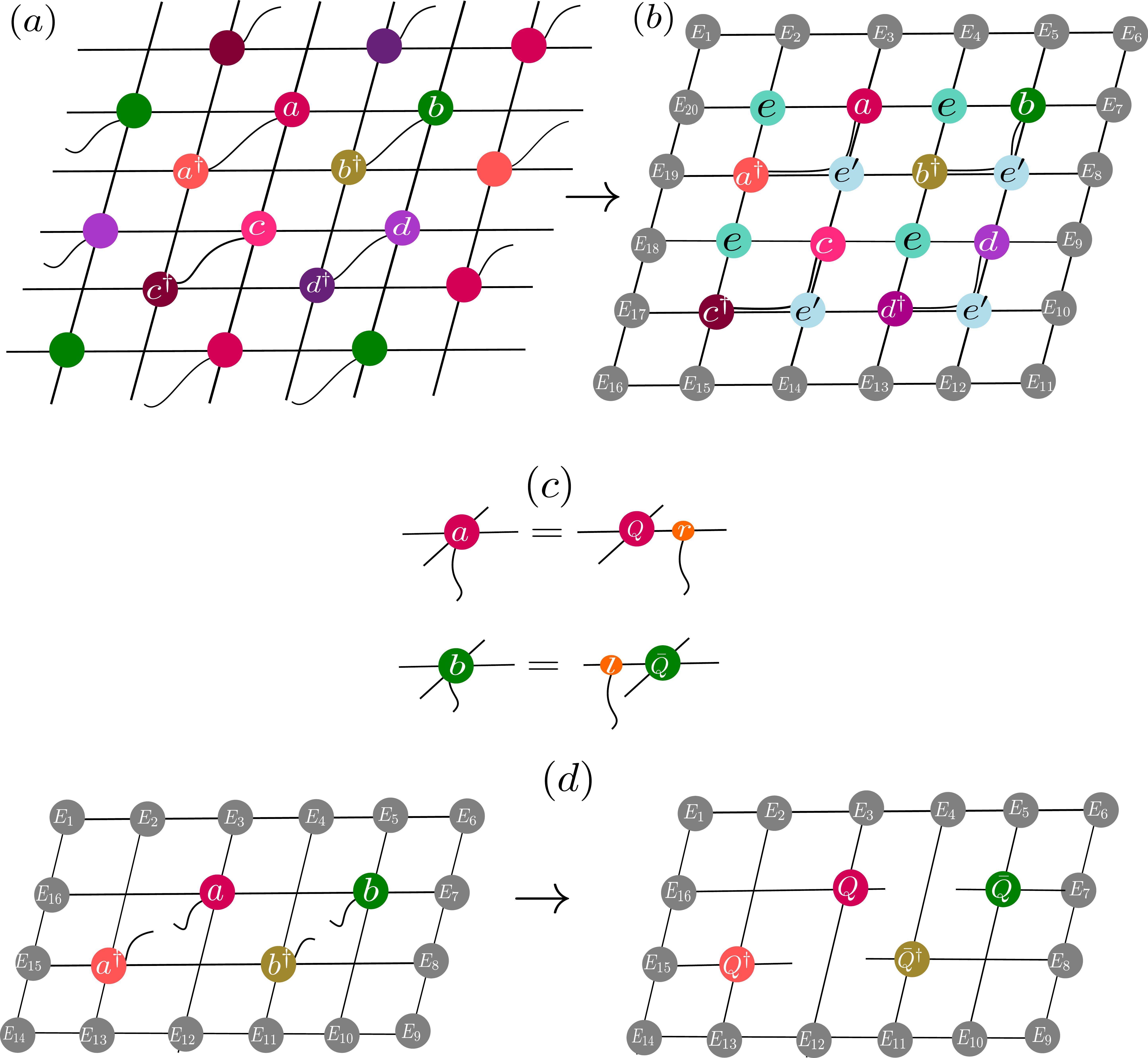} 
  \caption{(Color online) (a) The one-layer tensor network representation of the scalar product $\langle \psi| \psi \rangle$, where $| \psi \rangle$ represent a $2 \times 2 $ unit cell iPEPS, made of tensors $\{a, b, c, d\}$. (b) To approximate that, we need to use a standard CTM approach adapted to a $4 \times 4$ unit cell, made from tensors $\{a, b, c, d, a^{\dagger}, b^{\dagger}, c^{\dagger}, d^{\dagger}, e, e'\}$. (c) Tensor-network diagram of reduced-tensor application: Tensors $\{a, b\}$ are decomposed to low-rank tensors $\{l, r, Q, Q'\}$ by using LQ and QR decomposition. (d) Tensor-network representation of norm tensor $\mathcal{N}$. Right panel is obtained from left one by using reduced-tensor application and removing tensors $\{l, r, l^{\dagger}, r^{\dagger} \}$. The norm tensor $\mathcal{N}$ obtained by contracting whole right network, so that the optimal computation time scales like $\mathcal{O}(D^{9})$, achieved by a left-to-right step-by-step contraction.}
  \label{fig:ipeps1}
\end{center}
\end{figure}
 %%%%%%%%%%%%%%%%%%%%%%%%%%%%%%%%%%%%%

We assume the iPEPS $| \psi_{i} \rangle$ is made of a $2 \times 2$ unit cell including tensors $\{a, b, c, d\}$ and also assume that, e.g., interaction $e^{- \tau h_r}$ acts on tensors $\{a, b\}$. We need to find new tensors $\{a', b'\}$ to minimize the following cost function
\begin{equation*}
\min_{\{a',b'\}} \, f(| \psi_{i+1}(a',b') \rangle, e^{-\tau h_r} |\psi_{i}(a, b)\rangle),
\end{equation*}
where tensors $\{a',b'\}$ are considered as variational parameters. The cost function $f$ which stands for the square distance defined by
\begin{eqnarray*}
f=\langle \psi_{i}| {e^{- \tau h_r}}^{\dagger} e^{- \tau h_r}  |\psi_{i} \rangle+\langle \psi_{i+1} | \psi_{i+1}\rangle &\\- \langle \psi_{i+1}| e^{- \tau h_r} | \psi_{i} \rangle- \langle \psi_{i} |{e^{- \tau h_r}}^{\dagger}|\psi_{i+1}\rangle. 
\end{eqnarray*}
We use a a single-layer tensor network to represent the terms like $\langle \psi_{i+1} | \psi_{i+1}\rangle$ as shown in Fig.~\ref{fig:ipeps1}(a). They could then be approximated by a CTM approach adapted to $4 \times 4$ unit cells, as depicted in Fig.~\ref{fig:ipeps1}(b). To reduce computational cost of the optimization algorithm, we use a reduced-tensor scheme. The tensors $\{a',b'\}$ are split to subtensors $\{l,r,Q,\bar{Q}\}$ using QR and LQ decomposition, see Fig.~\ref{fig:ipeps1}(c). We rewrite the cost function as follows
\begin{equation*}
\min_{\{r, l\}} \, f= const+r^{\dagger}  l^{\dagger} \mathcal{N}  l  r - r^{\dagger}  l^{\dagger} \mathcal{N}-\mathcal{N}^{\dagger}r l,
\label{EQ:gaugefixing}
\end{equation*}
where $\mathcal{N}$ is called `norm tensor' as shown Fig.~\ref{fig:ipeps1}(d)---note that the first term does not play any role in the optimization procedure. We explicitly eliminate the negative part of the norm tensor $\mathcal{N}$ by replacing $\mathcal{N}$ by $\mathcal{N_{+}}$ in the cost function: $\mathcal{N_{+}}=\sqrt{\mathcal{\overline{N}}^{2}}$, where $\mathcal{\overline{N}}=(\mathcal{N}+\mathcal{N}^{\dagger})/2$. We then use an iterative way to minimize the cost function: we minimize the cost function with respect to $l$ by solving equation $\partial_{l^{\dagger}} f=0$ by holding fixed tensor $r$. Then we repeat this procedure for tensor $r$ with holding fixed $l$ until cost function converges. A careful analysis of the computational cost shows that, for all steps explained in this section, it scales as $\mathcal{O}(D^{9})$ . We find that this single-layer full-update simulation is quite robust providing the same accuracy and same advantage as double-layer one.

\begin{figure}[t]
\includegraphics[width = 1.0\linewidth]{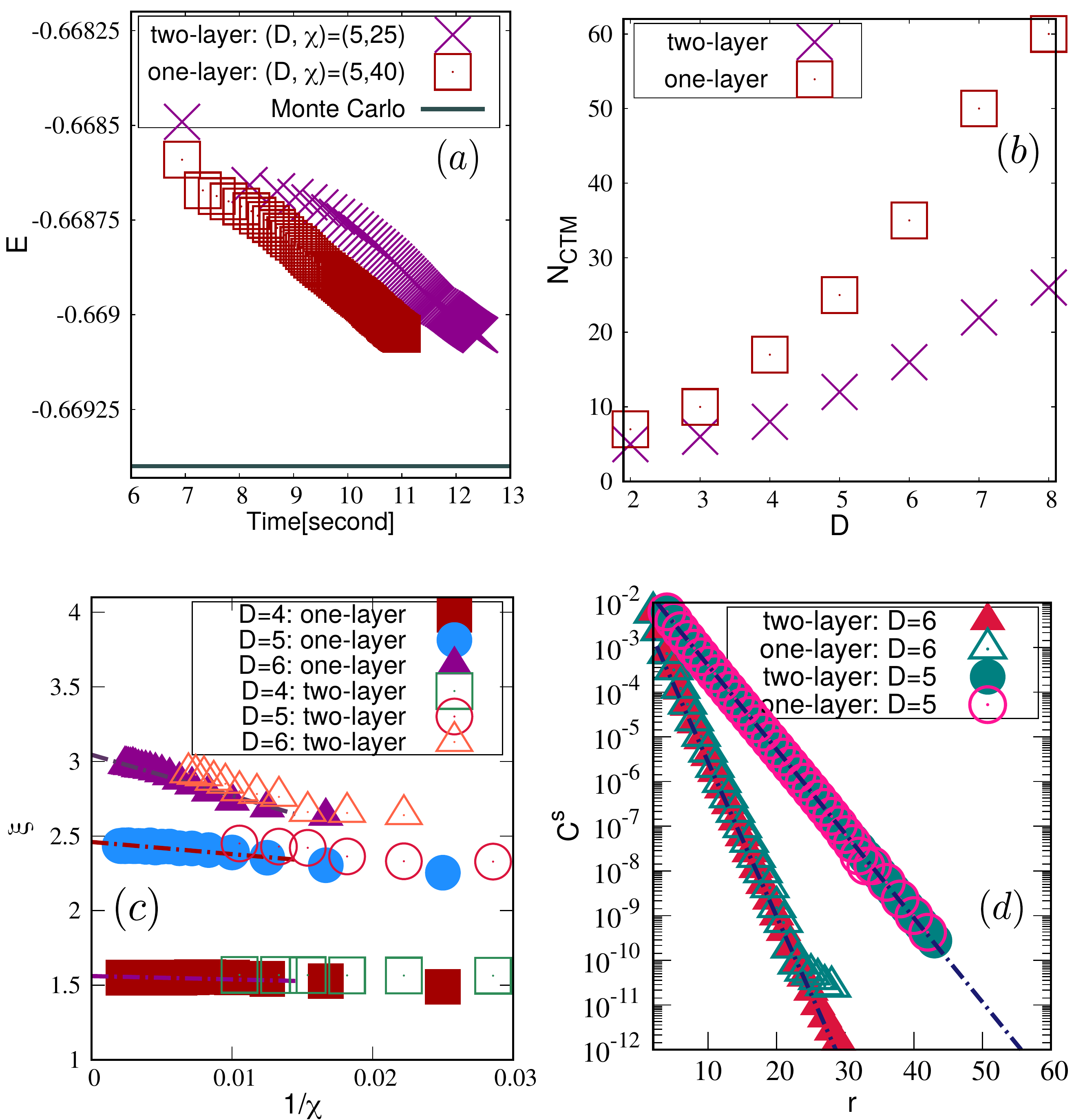}
\caption{ (Color online) A comparison between single- and two-layer tensor-network ansatz. (a) The variational energy $E$ as a function of computational time (seconds) in log scale for imaginary time $\tau=0.2$ and bond dimension $D=5$. (b) The iteration number in the CTM approach $N_{\text{CTM}}$ (to meet a threshold value) versus bond dimension $D$. (c) the characteristic correlation lengths $\xi$, obtained by transfer matrix, as a function of boundary bond dimension $\chi$ for both single-layer and two-layer tensor-network ansatz. (d) The Linear-log plot of spin-spin (only $z$ component) correlation function versus distant $r$.} 
\label{fig:energy}
\end{figure}

\subsection{Benchmark results}
\label{Sec:Benchmark}
We benchmark the single-layer full-update scheme by studying the Heisenberg model on an infinite square lattice. Our goal is to directly compare the single-layer scheme with double-layer one and show that the former provides the same level of accuracy for local and non-local quantities (similar to double-layer scheme) with much less computational time. Thus, one might choose the single-layer schemes (as a more efficient toolkit) over double-layer ones. We study its stability and to compare its accuracy with that of previous double-layer ones---say, double-layer full-update scheme. The model is defined by 
\begin{equation*}
H=\sum_{\langle i,j\rangle}\textbf{S}_{i}\cdot\textbf{S}_{j},
\end{equation*}
where ${\textbf{S}_i}\equiv(\textbf{S}_i^x, \textbf{S}_i^y, \textbf{S}_i^z)$ are spin-$1/2$ operators---the sum goes over the first-neighbor sites. We use a $2 \times 2$ unit-cell iPEPS ansatz, made of tensors $\{a, b, c, d\}$, to obtain ground-state wave function of the Heisenberg model. We have explicitly used the same algorithm explained in Sec.~\ref{Sec:single-layer-full-update} to perform the optimization algorithms. The observables are obtained by a modified CTM renormalization group approach explained in Refs.~\onlinecite{Corboz:2014, Huang:2012}. In this study, we do not enforce any symmetry on the tensors, filling them with total random complex numbers. The maximum bond dimensions that we could afford for a single-layer iPEPS ansatz with simple-update simulation and full-update simulation are respectively $(D, \chi)=(10, 100)$ and $(D, \chi)=(16, 400)$ on square or Kagome lattice.

In Fig.~\ref{fig:energy}(a), we have compared the variational energy obtained by a single-layer and double-layer full-update simulations versus computational time for imaginary time $\tau=0.2$ and bond dimension $D=5$. We have used the precise Monte Carlo energy $-0.6694$~\cite{Sandvik:1997} as the reference energy. In both cases, while starting from the same initial iPEPS state, the variational energy converges to the same value at the end, providing the same accuracy. But in single-layer simulation, the convergence rate (to reach a specific value) is remarkably faster, as expected due to the cheaper computational cost of the algorithm. We also observe that, when $\tau \rightarrow 0$, both algorithms provide the same accuracy as there is no difference in the final variational ground-state energy: for example for bond dimension $(D, \chi_{\text{one-layer}}, \chi_{\text{two-layer}})=(6,230,120)$, the final variational ground-state energy would be $E=-0.66932$ for both methods---which is close to Monte Carlo result. This observation is also valid for other bond dimensions.

We notice that, although, leading computational cost of the single-layer algorithm is smaller, but the prefactor (even in sub-leading computational costs) could play a key role in practical running times. For example, as shown in Fig.~\ref{fig:energy}(a), the iteration number $N_{\text{CTM}}$ (to obtain converged results) in the CTM approach is much larger in the single-layer case. This is understandable as in a single-layer picture, we need to treat $4 \times 4$ unit cells, which requires more iteration number to converge than a $2 \times 2$ unit cell in the double layer. This factor turns out to be $2, 3$ for this case and roughly remains the same for all bond dimensions. In addition, a larger boundary dimension is required for single-layer algorithm to provide the same accuracy as double-layer one~\cite{Xie:2017}, as empirically we find to have the same level of accuracy one should choose $\chi_{\text{sinlge}} \sim 2 \chi_{\text{double}}$. 

%%%%%%%%%%%%%%Fig. 3%%%%%%%%%%%%%%%%%%
\begin{figure}
\begin{center}
\includegraphics[width=1.0 \linewidth]{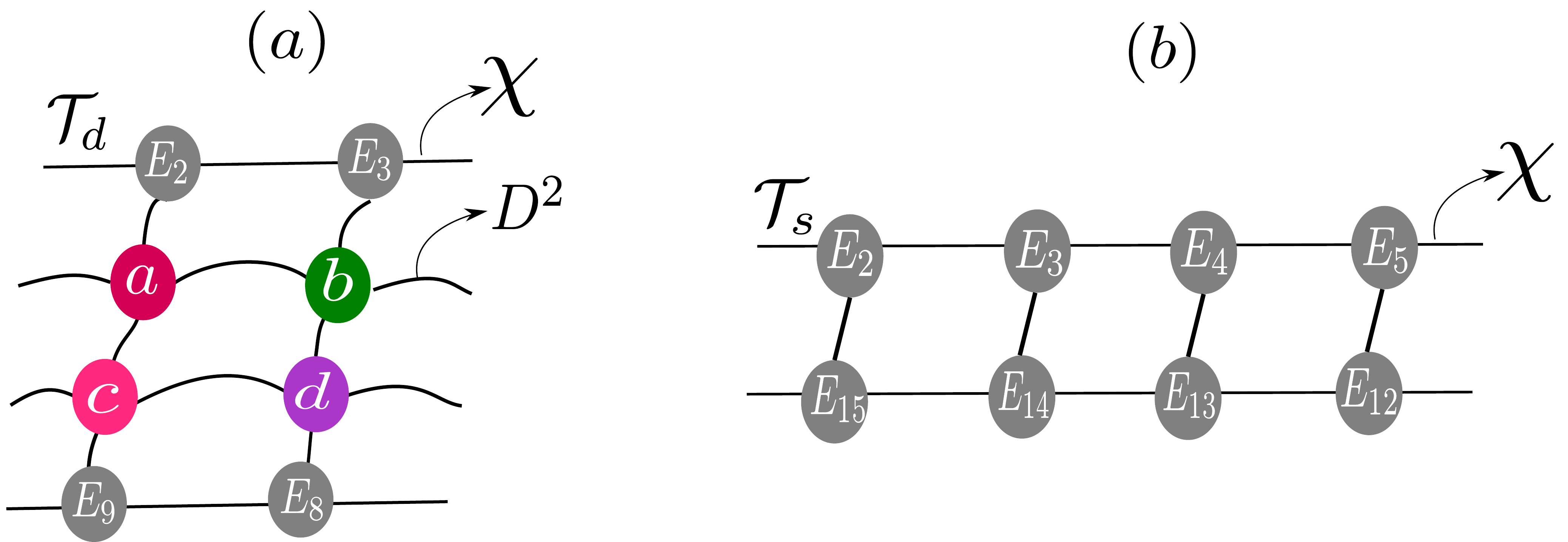} 
  \caption{(Color online) Tensor-network representation of transfer matrix. $(a)$ $\mathcal{T}_{d}$ and $(b)$ $\mathcal{T}_{d}$ provide two diffident representations to estimate correlation length of the system---with diffident computational cost and accuracy. In the $\chi \rightarrow \infty$ limit, estimated correlation lengths should converge to the same value $\xi_{d} \rightarrow \xi_{s}$. $\mathcal{T}_{d}$ and $\mathcal{T}_{s}$ are respectively used in double-layer and single-layer tensor-network ansatz.}
  \label{fig:TM}
\end{center}
\end{figure}
 %%%%%%%%%%%%%%%%%%%%%%%%%%%%%%%%%%%

We also study correlation function to see how a single-layer algorithm could capture the non-local properties of the system. Specifically, we study the correlation length $\xi$ obtained by the row-to-row transfer matrix $\mathcal{T}$ as depicted in Fig.~\ref{fig:TM}(a, b) and also spin-spin correlation function $C_s(r)$ as a function of distance $r$. As a side remark, we notice that two different representations of transfer matrix $\mathcal{T}$ is possible to be used to estimate the correlation length $\xi$: one is only made of environment tensors, denoted by $\mathcal{T}_{s}$, and other one which also includes local tensors $\{a, ,b, c, d\}$, denoted by $\mathcal{T}_{d}$. In the $\chi \rightarrow \infty$ limit, one expects the correlation lengths, extracted from different representations, to converge to the same value, $\xi_{d} \rightarrow \xi_{s}$ (as environment tensors truly represent the infinite column of tensors). However, for a fixed value of $\chi$, one expects that $\mathcal{T}_{d}$ would estimate more accurately correlation length than $\mathcal{T}_{s}$---as it provides a more accurate representation of an infinite row of tensors. Although the representation of $\mathcal{T}_{s}$ has this advantage that it requires cheaper computational cost, i.e. $\chi^{5}$, to obtain correlation length---by using an eigensolver library like Arnoldi. The leading eigenvalues $\lambda_i$ of the transfer matrix provide an estimation of correlation length, as $\xi=\frac{-1}{\log(|\frac{\lambda_{2}}{\lambda_{1}}|)}$ where $\lambda_{1}$ and $\lambda_{2}$ ($|\lambda_{1}|>|\lambda_{2}|$) are respectively the largest eigenvalues.

In Fig.~\ref{fig:energy}(c), we have plotted the correlation lengths $\xi$ as a function of boundary bond dimension $\chi$, obtained from simulations of single-layer and double-layer full-update scheme for Heisenberg model. We respectively utilize representations of $\mathcal{T}_{s}$ and $\mathcal{T}_{d}$ for single-layer and double-layer algorithms to extract correlation length. It seems that both methods predict the same value of the correlation lengths in the $\chi \rightarrow  \infty$ limit, although larger boundary bond dimensions are required in the single-layer algorithm to do a reliable scaling. The same consistency is observed in spin-spin correlation function, shown in Fig.~\ref{fig:energy}(d). It implies that a single-layer algorithm could faithfully capture also long-range behavior of the system similar to double-layer one.

%%%%%%%%%%%%%%%%%%%%%%%%%%%%%%%%%%%%%%%%%%%%%%%%%%%%
\section{Antiferromagnetic Heisenberg model with scalar-chiral interaction} 
\label{sec:chiral}

The second model we consider is the spin-$1/2$ kagome-lattice Heisenberg model with scalar-chiral interaction, which is defined as
\begin{equation*}
H=\sum_{\langle i,j\rangle} \textbf{S}_{i}\cdot\textbf{S}_{j}  + J_{ch}  \sum_{i, j, k \in \bigtriangleup} \textbf{S}_{i}   \cdot (\textbf{S}_{j} \times \textbf{S}_{k}),
\end{equation*}
where for the three-spin scalar-chiral interaction $J_{ch} \textbf{S}_{i}   \cdot (\textbf{S}_{j} \times \textbf{S}_{k})$ the sum goes over all elementary triangles of the kagome lattice. The three spins $i,j,k$ for all the triangles follow the clockwise direction. This chiral term breaks time-reversal symmetry and parity symmetry. In the pure Heisenberg model with $J_{ch}=0$, the ground state has been identified as a spin liquid state. The recent promising studies mostly support either a gapped $\mathcal{Z}_2$ spin liquid~\cite{Yan:2011, jiang2012, depenbrock2012} or a gapless $U(1)$ Dirac spin liquid~\cite{Liao:2017, He:2017} as the ground-state candidate. Here we refer to this spin liquid of the Heisenberg model as kagome spin liquid.

In the other extreme case $J_{ch}=\infty$ (equivalent to the model with only the chiral interaction term), a recent DMRG study~\cite{Bauer:2014} identifies the phase to be a gapped chiral spin liquid, inheriting the universal features of the $\nu=1/2$ Laughlin state including the chiral edge mode and the modular matrix that describes the statistic properties of the quasiparticles. In addition, based on the ground-state fidelity, an upper bound of quantum phase transition point is predicted at $J_{ch} \simeq 0.16$ (we take the Heisenberg coupling as the energy scale), where the growing $J_{ch}$ coupling drives the kagome spin liquid to the gapped chiral spin liquid~\cite{Bauer:2014}. Our main goal is to study the bulk properties of this gapped phase by using a single-layer iPEPS ansatz to investigate whether the iPEPS state could faithfully represent this chiral spin liquid. Furthermore, we estimate the quantum phase transition between the two spin liquid phases by studying the derivative of the chiral order parameter. We also compare our iPEPS results with those from DMRG calculation.

\subsection{Local order parameters } 
\label{sec:Accuracy}

\begin{figure}[ht]
  \centering
 \includegraphics[width = 1.0\linewidth]{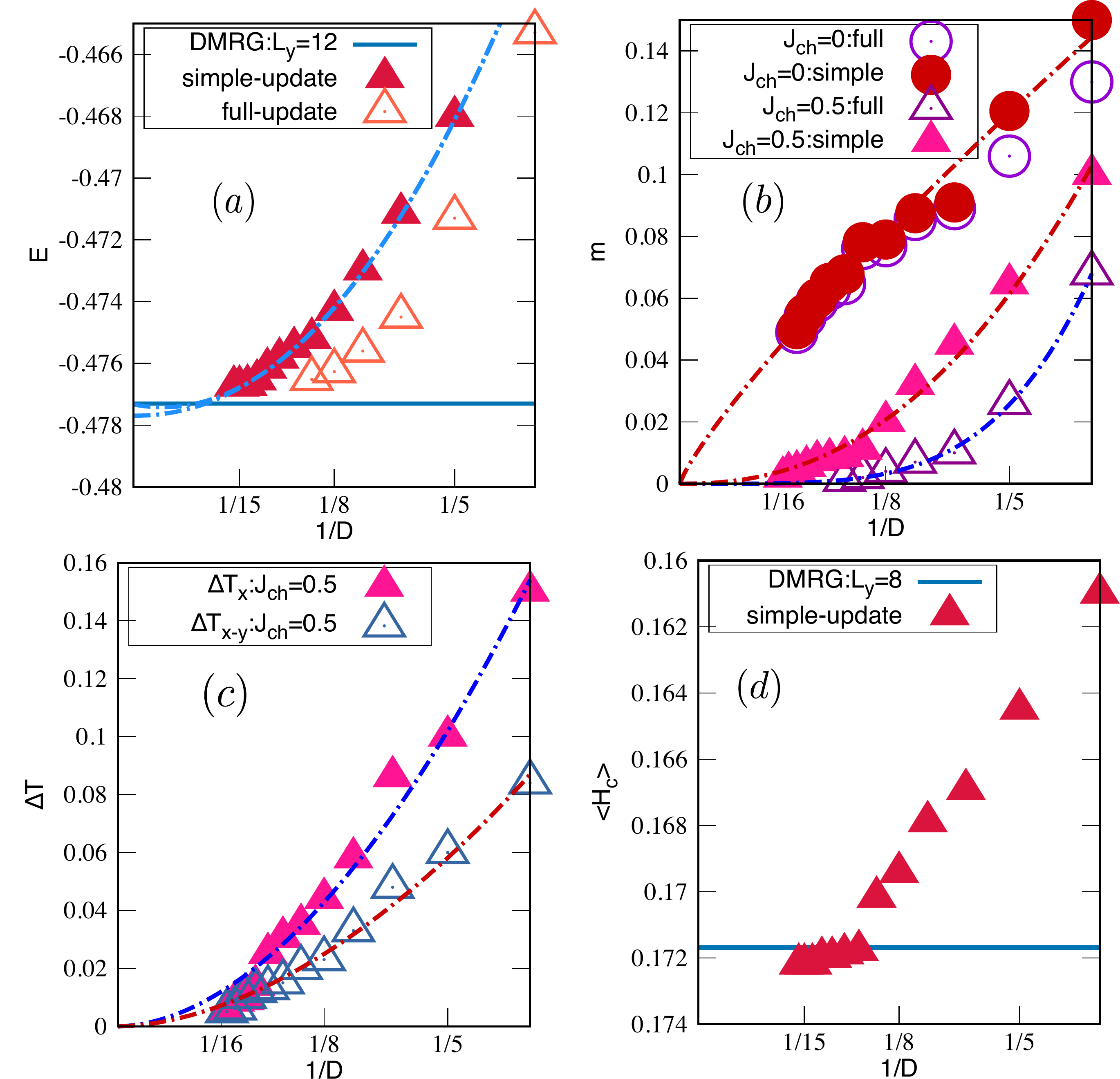}
 \caption{ (Color online) (a) The iPEPS variational ground-state energy $E$, obtained by using full and simple update in a single-layer tensor-network framework, versus bond dimension $1/D$ at point $J_{ch}=0.5$. DMRG result is for a cylinder with width $L_y=12$. (b) The magnetic order parameter $m$ as a function of bond dimension $1/D$ at points $J_{ch}=0, 0.5$. The dashed lines stands for an algebraic fit $m \sim D^{-\beta}$. Our estimation of the exponent $\beta$ is $0.6$ and $5$ respectively for points $J_{ch}=0$ and $J_{ch}=0.5$ by using full update. (c) The lattice symmetry-breaking parameter $\Delta T$ and (d) the scalar chirality term $H_{c}$ as a function of bond dimension $1/D$.}
 \label{fig:Energy5}
\end{figure}

First of all, we study the variational ground-state energy $E$, magnetic order parameter $m$, and local bond energy difference $\Delta T$ to investigate the accuracy of the single-layer iPEPS ansatz and to study the nature of the system\cite{note:iPEPS}. In Fig.~\ref{fig:Energy5}(a), we show the variational ground-state energy $E$ as a function of bond dimension $1/D$ at $J_{ch} = 0.5$ (for this point the system is in the chiral spin liquid phase as predicted by DMRG study~\cite{Bauer:2014}). It can be seen that full-update simulation obtains the lower energy compared to simple-update for a fixed bond dimension $D$. Nonetheless, we could always get the optimized variational energy by using simple-update simulation as the larger bond dimension $D$ could be accessed. In addition, we find that simple-update simulation never gets stuck in local minimum as the energy decreases monotonically in our calculation (also see Ref.~\onlinecite{Liao:2017} for the similar discussion). By using a polynomial fitting, the ground-state energy in the thermodynamic limit is estimated as $E_{\text{iPEPS}} \simeq -0.4775$ in the $D\rightarrow \infty$ limit. As discussed in Ref.~\onlinecite{Liao:2017} at $J_{ch}=0$, the authors find an algebraic  scaling behavior of ground-state energy as $E \sim e_{0}+b \times D^{-\alpha}$. 
Here we could still faithfully fit our data using the same function, as shown by the blue dashed lines in Fig.~\ref{fig:Energy5}(a). The exponents $\alpha$ for simple- and full-update are nearly $2$ and $3$, respectively. Here we remark that with our current data points, it would be difficult to say that the energy could only be scaled algebraically as other types of polynomial fittings could also be valid. For a benchmark, we also show the energy obtained by $SU(2)$-symmetric DMRG on a cylinder with the width $L_y = 12$ ($L_y$ is the number of sites along the circumference of cylinder). The DMRG results converge fast with cylinder circumference (as expected for a gapped state), which is $E_{\text{DMRG}} \simeq -0.4773$ for $L_{y}=12$. We can see the good agreement between DMRG and single-layer iPEPS ansatz, demonstrating the validity of our iPEPS results.

Next we calculate the magnetic order parameter $m = \sqrt{\langle S_x \rangle^2 + \langle S_y \rangle^2 + \langle S_z \rangle^2}$ for different $J_{ch}$. As shown in Fig.~\ref{fig:Energy5}(b), the magnetic order $m$ for $J_{ch} = 0.5$ decreases rapidly with increasing bond dimension $D$, reaching the value of about $10^{-3}$ for the largest bond dimension $D=16$. We also notice that for full-update simulation, $m$ drops off faster than simple-update simulation, which seems to vanish even for a finite bond dimension. The vanishing magnetic order parameter indicates the absent magnetic order in this phase. We also compare these results with the one of the kagome spin liquid with $J_{ch}=0$\cite{Kagome:2017}. As discussed in Ref.~\onlinecite{Liao:2017}, the magnetic order parameter $m$ at $J_{ch}=0$ shows an algebraic fall-off with respect to bond dimension $D$, i.e. $m \sim D^{-\beta}$ with $\beta \simeq 0.6$, which has also been observed in our calculation as shown in Fig.~\ref{fig:Energy5}(b). This algebraic vanishing of the magnetization has been suggested as evidence of a gapless spin liquid in Ref.~\onlinecite{Liao:2017}. For $J_{ch} = 0.5$ we find that  the scaling behavior of $m$ could also be fitted as $m \sim D^{-\beta}$ with $\beta =  3$ and $5$ for simple-update and full-update simulations, respectively. On the other hand, these extrapolation behaviors of $m$ seem more reasonable to be fitted by an exponential decay, especially for the largest bond dimensions, which may be related to the gapped nature of the chiral spin-liquid phase.

We also calculate bond energy difference $\Delta T$ to study translational symmetry breaking: $\Delta T_{x}=\max(E_{x})-\min(E_{x})$, and $\Delta T_{x-y} = \max(E_{y})-\min(E_{x})$. The symbols $E_x$ and $E_y$ denote the bond energy for the horizontal and vertical directions in the unit cell. A non-zero value of either $\Delta T_{x}$ or $\Delta T_{x-y}$ in the $D \rightarrow \infty$ limit represents a spontaneous lattice symmetry breaking. In Fig.~\ref{fig:Energy5}(c), we demonstrate these bond energy differences as a function of $1/D$, where both $\Delta T_x$ and $\Delta T_{x-y}$ are vanishing-small ($\sim 10^{-4}$) in the large-$D$ limit. These results indicate the absence of a lattice rotational/translational symmetry breaking, which is in agreement with the chiral spin-liquid phase found by DMRG~\cite{Bauer:2014}.

%%%%%%%%%%%%%%%%%%%%%%Table%%%%%%%%%%%%%%%%%%%%%%%%%%
\begin{table}
\begin{tabular}{|c|c|c|c|c|}
\hline $J_{ch}$ &$E_{\text{iPEPS}}$ &$m$ &$\Delta T_{x-y}$&$\langle H_c \rangle$ \\ \hline
$J_{ch}=0.5$ &$-0.4768$&$0.004$ &$0.004$&$0.172$ \\
$J_{ch}=2.0$ &$-0.6889$&$0.006$&$0.002$&$0.231$ \\
$J_{ch}=\infty$ &$-0.1715$ & $0.009$&$0.0005$&$0.254$ \\  \hline
\end{tabular}
\caption{The iPEPS results for diffident coupling parameters $J_{ch}$ and bond dimension $D=14$.}
\label{tab:results}
\end{table}
%%%%%%%%%%%%%%%%%%%%%%%%%%%%%%%%%%%%%%%%%%%%%%%%%%%%%%

In addition, we present the results of the scalar chiral order $\langle H_{c} \rangle =\langle \textbf{S}_{i}  \cdot (\textbf{S}_{j} \times \textbf{S}_{k}) \rangle$ as a function of bond dimension $1/D$. As depicted in Fig.~\ref{fig:Energy5}(d), the expectation value of $\langle H_{c}\rangle$ (averaged on all local triangles) converges quite fast, especially for large bond dimensions $D>10$. The chiral order is estimated to be $0.1721$ in the $D \rightarrow \infty$ limit, which is quite close to the DMRG result $0.1717$ obtained on a cylinder with width $L_y=8$.

We observe almost the same behaviors of these local order parameters as a function of bond dimension $D$ for $J_{ch} \geq 0.5$. In Table.~\ref{tab:results}, we show some results at $J_{ch} = 0.5, 2.0$ and $J_{ch} = \infty$ with bond dimension $D=14$ for further reference.

\begin{figure}[ht]
\centering
\includegraphics[width = 1.0\linewidth]{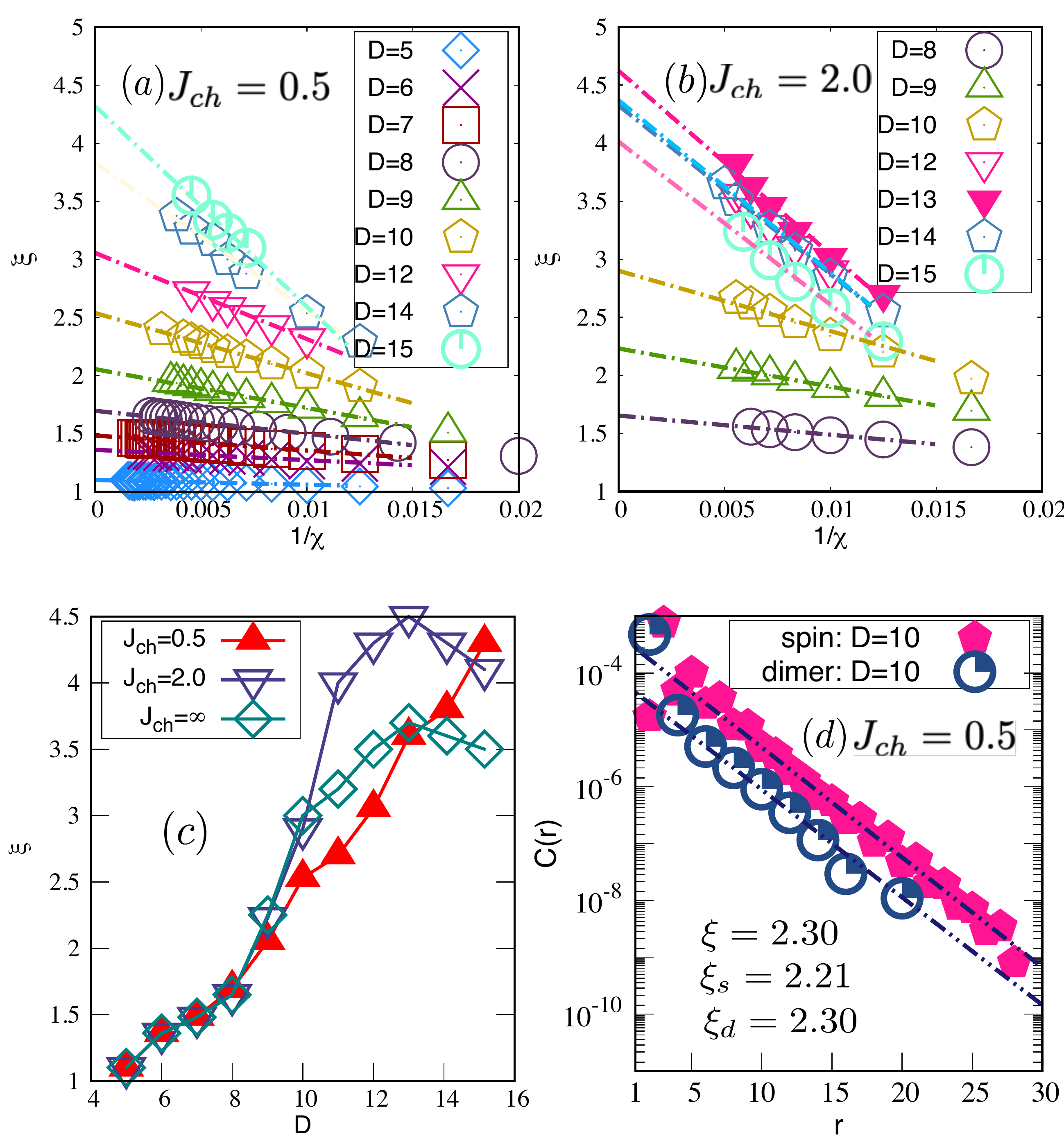}
\caption{ (Color online) (a)-(b) The correlation length $\xi$, extracted from transfer matrix $\mathcal{T}_s$, as a function of boundary bond dimension $\chi$ for $J_{ch}=2, 0.5$. (c) The correlation length $\xi$ versus bond dimension $D$ for $J_{ch} = 0.5, 2, \infty$. Here $\xi$ has been extrapolated to the infinite-$\chi$ limit. (d) The linear-log plot of spin-spin and dimer-dimer correlation functions versus distance $r$ for $J_{ch} = 0.5$. The slopes show inverse of the associated spin and dimer correlation length $\xi^{-1}_{s}$ and $\xi^{-1}_{d}$.}
\label{fig:EnergyK1}
\end{figure}

\subsection{Correlation length}
\label{sec:Correlation-length}
To further characterize the chiral spin-liquid phase, we compute the correlation length $\xi$ of the system by calculating the leading eigenvalue of the transfer matrix $\mathcal{T}_{s}$~\cite{Poilblanc:2018, Corboz:2018} as sketched in Fig.~\ref{fig:TM}(b).
We expect a short correlation length in gapped chiral spin-liquid phase. 
We focus on the phase region with $0.3 < J_{ch} \leq \infty$. 
Since the correlation length could be computed by using a row-to-row or a column-to-column transfer matrix, here we only report the maximum value. First of all, we investigate the boundary bond dimension ($\chi$) dependence of correlation length, which are shown in Fig.~\ref{fig:EnergyK1}(a-b). 
One can see that correlation length has a strong dependence on $\chi$ in the calculations with a large bond dimension $D$.
For each given bond dimension $D$, we estimate the correlation length in the $\chi \rightarrow \infty$ limit by using a linear fitting on the largest values of $\chi$, which always leads to a finite correlation length in the infinite-$\chi$ limit for our studied bond dimensions.
Interestingly, with growing bond dimension $D$, the correlation length extrapolated in the infinite-$\chi$ limit displays different behaviors.
For $J_{ch} = 0.5$, $\xi$ monotonically increases with growing $D$, but for $J_{ch} \geq 2.0$ correlation length reaches a fixed value or starts to decrease slightly for large bond dimension, which are shown in Fig.~\ref{fig:EnergyK1}(c).
For $J_{ch} = 0.5$ and $\infty$, the correlation length that seems to be finite in the infinite-$D$ limit is consistent with the nature of a gapped spin liquid. Furthermore, our results indicate that although the system has a short correlation length, a relatively large bond dimension may be required to demonstrate the finite correlation length in our present iPEPS ansatz. 
At the smaller $J_{ch} = 0.5$, correlation length keeps growing in our studied bond dimension. 
Since this point is still in the gapped chiral spin-liquid phase, we believe that a finite correlation length might be observed by keeping larger bond dimension, which is however beyond the capability of our current computation.

To get more insight into the behavior of correlation length, we also study the correlation functions. 
We extract the correlation lengths associated with physical operators and compare them with the one obtained from the transfer matrix. 
As shown in Fig.~\ref{fig:EnergyK1}(d) for $J_{ch} = 0.5$, we plot spin-spin and dimer-dimer correlation functions $C(r)$ versus site and bond distance $r$ in the logarithmic scale. 
A finite bond dimension $D$ presumably induces an exponential decay of correlation function for long distance, i.e. $C(r) \sim e^{-\frac{r}{\xi}}$.
Thus, we could use the formula $\log(C(r)) = -\frac{r}{\xi} + const$ for large $r$ to extract the associated correlation length $\xi$. 
We find that both spin ($\xi_s$) and dimer ($\xi_d$) correlation lengths are quite close to the one obtained from the transfer matrix.  
For $J_{ch} > 0.5$ we observe a similar behavior, which we do not show here.

\begin{figure}[h]
  \centering
\includegraphics[width = 1.0\linewidth]{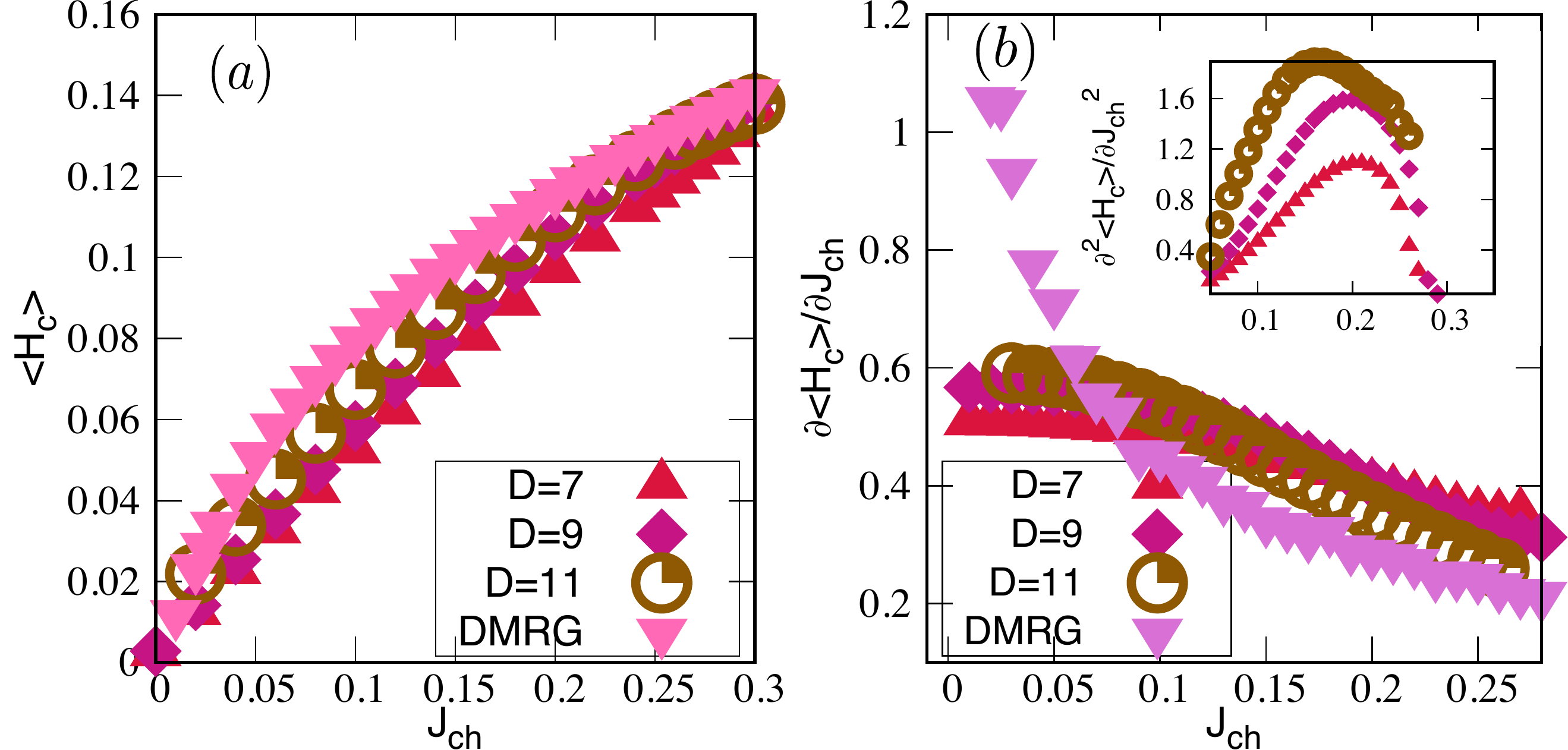}
\caption{(Color online) The scalar chiral order $\langle H_c \rangle$ and its derivative as a function of coupling parameter $J_{ch}$ for different bond dimensions. The DMRG results are obtained on the $L_y = 8$ cylinder system.}
\label{fig:chiral}
\end{figure}

\subsection{phase diagram} 
\label{sec:phase}

To establish the phase diagram of the model, in which the kagome spin liquid is separated from the chiral spin liquid, we calculate the derivative of the chiral order $\langle H_c \rangle$ ($\langle H_c \rangle = \langle \textbf{S}_{i}   \cdot (\textbf{S}_{j} \times \textbf{S}_{k}) \rangle$) with respect to $J_{ch}$. We first compare the chiral order $\langle H_c \rangle$ obtained by iPEPS with the DMRG results (obtained on the $L_y = 8$ cylinder) to check the validity of our data. As seen in Fig.~\ref{fig:chiral}(a), it seems that both data are in good agreement for $J_{ch} > 0.25$ although DMRG results are obtained on finite-size system. The first derivative of $\langle H_c \rangle$ with respect to $J_{ch}$ has been shown for diffident bond dimensions in Fig.~\ref{fig:chiral}(b). For iPEPS results, we do not observe a singularity in the first derivative $\partial \langle H_c \rangle / \partial J_{ch}$ by increasing bond dimension. Interestingly, the second derivative $\partial^2 \langle H_c \rangle / \partial J_{ch}^2$ shows different behavior, see Fig.~\ref{fig:chiral}(b, inset): it increases with growing $J_{ch}$ and goes through a peak around $J_{ch} \simeq 0.17$ (for bond dimension $D=11$), and it drops off for larger $J_{ch}$\cite{DMRG:note}. This peak which shifts with increasing bond dimension may indicate a quantum phase transition from the kagome spin liquid to the chiral spin liquid. A finite-$D$ scaling of the peak is required to accurately determine the transition point in the $D \rightarrow \infty$ limit. These iPEPS data may be consistent with a transition around  $J_{ch} \approx 0.14$, while a similar transition region (transition $J_{ch} < 0.16$) was estimated by computing wavefunction overlap in previous DMRG calculation~\cite{Bauer:2014}. However, we emphasize that in current SU(2)  DMRG calculations, the first derivative $\partial \langle H_c \rangle / \partial J_{ch}$ clearly deviates from the iPEPS results for $J_{ch} \lesssim 0.1$, which grows rapidly with decreasing $J_{ch}$, showing a strong response of the kagome spin liquid to the chiral interaction.  From these results,  we conjecture that the critical $J_{ch}$ may be quite small based on our DMRG data, as the quantum system may enter the gapped chiral state from a gapless state for very small $J_{ch} \sim 0.01$. iPEPS results by keeping larger bond dimension and DMRG data on larger system sizes are required in future study to pin down the chiral response of the kagome spin liquid for weak chiral interactions.

%%%%%%%%%%%%%%%%%%%%%%%%%%%%%%%%
\section{Summary and discussion}
\label{Sec:CONCLUSION}
In this paper, we have discussed in detail the implementation of iPEPS ansatz in the framework of the single-layer tensor network by using corner transfer matrix method~\cite{SL:note, Xie:2017}. It is shown that all the steps could be accomplished with a computational cost $\mathcal{O}(D^9)$ ($\mathcal{O}(D^{12})$ in the double-layer algorithm). By studying the square Heisenberg model, we provide the benchmark data to show the stability and accuracy of this single-layer algorithm and compare it with the previous double-layer-based method. We find that the single-layer algorithm is quite stable and provides the same level of accuracy as the double-layer method. Additionally, by studying correlation length and correlation function, we observe it truly captures long-range behavior of the system, similar to the double-layer algorithm. Our study shows that \emph{only disadvantage} of this single-layer scheme is that it requires larger boundary bond dimension and larger number of iterations for CTM (as we are dealing with larger unit cells) to reach the same level of accuracy as double-layer one.

By using our single-layer iPEPS ansatz, we investigate the bulk properties of the kagome Heisenberg model with additional scalar-chiral interaction. We systematically study the bond-dimension ($D$) scaling of magnetization and local bond energy difference (measuring lattice symmetry breaking) in the chiral spin liquid phase. It is observed that both order parameters rapidly go to zero in the $D\rightarrow \infty$ limit, supporting the chiral spin liquid identified by DMRG~\cite{Bauer:2014}. In order to examine the gap nature of this spin liquid in our iPEPS representation, we calculate the correlation length extracted from row-to-row transfer matrix. We observe that for strong chiral interactions $J_{ch} > 2$, correlation length increases with growing bond dimension $D$ and finally reaches a fixed value for the largest bond dimensions $D > 13$. It indicates the gapped nature of the chiral spin-liquid phase, which is consistent with DMRG results. In addition, this observation implies that iPEPS ansatz may require large bond dimension to faithfully capture the gap nature of a gapped chiral spin liquid. For small bond dimension, correlation length $\xi$ grows with the behavior like $\xi \sim D^{1.1}$. Simulations with small bond dimension may lead to the wrong conclusion. The finite correlation length identified in our iPEPS simulation for the gapped chiral spin liquid may shed new light on further PEPS construction of gapped chiral topological states. Furthermore, we study the derivative of the chiral order and we estimate the quantum phase transition between the kagome spin liquid and the chiral spin liquid. While a critical  $J_{ch} \simeq 0.14$ is found based on our studied bond dimensions of iPEPS, which may be an upper bound of the transition point entering
gapped chiral phase. For small $J_{ch}$ ($J_{ch} \lesssim 0.1$), the response of the kagome spin liquid to the additional chiral term found in iPEPS simulation seems quite different from our DMRG results. This disagreement may come from the reason that for weak $J_{ch}$, a much larger bond dimension may be required to obtain well converged chiral order parameter.

Our study of a single-layer iPEPS ansatz could be significant progress toward addressing more challenging problems of frustrated magnetism. One can consider possible other avenues to further improve it. For example, a large bond dimension could be reached by simply implementing global symmetry into this scheme and better ground-state energy may be reached by using a variational ansatz. This scheme could promisingly shed light on the true nature of phases of interacting fermionic systems such as the $t$-$J$ and Hubbard models with competing orders.

\emph{Note: Upon finishing this work, we noticed a recent related work presented in Ref.~\onlinecite{Lee:2018}.}

%%%%%%%%%%%%%%%%
\acknowledgments
Work by RH and DNS was supported by the Department of Energy, Office of Basic Energy Sciences, Division of Materials Sciences and Engineering, under Contract No. DE-AC02-76SF00515 through SLAC National Accelerator Laboratory. 
The research was partly supported by the National Natural Science Foundation of China Grants 11834014, 11874078 (S.S.G.), and by the Fundamental Research Funds for the Central Universities (S.S.G.). We have used \emph{Uni10} \cite{Kao:2015} as a middleware library to build the iPEPS ansatz. 
%%%%%%%%%%%%%%%%%%%%%%

\bibliography{Ref}

\begin{thebibliography}{70}
\bibitem [{not()}]{note:iPEPS}%
  \BibitemOpen
  \href@noop {} {}\bibinfo {note} {We always use the outcome of iPEPS ansatz
  with the reduced bond dimension $D-1$ to initialize the one with bond
  dimension $D$. It might increase the chance of getting stuck in a local
  minimum, but we find this way reliable to obtain the best variational
  energy.}\BibitemShut {Stop}%
\bibitem [{DMR()}]{DMRG:note}%
  \BibitemOpen
  \href@noop {} {}\bibinfo {note} {We could not obtain smooth curve for the
  second derivative of the chiral order by using DMRG data.}\BibitemShut
  {Stop}%
\end{thebibliography}

\end{document}